\documentclass[conference]{IEEEtran}

\usepackage[latin9]{inputenc}
\usepackage{amsmath}
\usepackage{amssymb}
\usepackage{graphicx}
\usepackage{fancyhdr}
 
\DeclareTextSymbolDefault{\textquotedbl}{T1}

\IEEEoverridecommandlockouts
\usepackage{cite}
\usepackage{amsfonts}\usepackage{algorithmic}
\usepackage{subfigure}
\usepackage{textcomp}
\usepackage{xcolor}
\def\BibTeX{{\rm B\kern-.05em{\sc i\kern-.025em b}\kern-.08em
    T\kern-.1667em\lower.7ex\hbox{E}\kern-.125emX}}

\makeatother

\begin{document}
\title{Exceptional Point of Degeneracy in Backward-Wave Oscillator with Distributed
Power Extraction}
\author{\IEEEauthorblockN{Tarek Mealy, Ahmed F. Abdelshafy and Filippo Capolino}
\IEEEauthorblockA{\textit{Department of Electrical Engineering and Computer Science,
University of California, Irvine, CA 92697 USA} \\
 tmealy@uci.edu, abdelsha@uci.edu and f.capolino@uci.edu}}
\maketitle
\thispagestyle{fancy}
\begin{abstract}
We show how an exceptional point of degeneracy (EPD) is formed in
a system composed of an electron beam interacting with an electromagnetic
mode guided in a slow wave structure (SWS) with distributed power
extraction from the interaction zone. Based on this kind of EPD, a
new regime of operation is devised for backward wave oscillators (BWOs)
as a synchronous and degenerate regime between a backward electromagnetic
mode and the charge wave modulating the electron beam. Degenerate
synchronization under this EPD condition means that two complex modes
of the interactive system do not share just the wavenumber, but they
rather coalesce in both their wavenumbers and eigenvectors (polarization
states). In principle this new condition guarantees full synchronization
between the electromagnetic wave and the beam's charge wave for any
amount of output power extracted from the beam, setting the threshold
of this EPD-BWO to any arbitrary, desired, value. Indeed, we show
that the presence of distributed radiation in the SWS results in having
high-threshold electron-beam current to start oscillations which implies
higher power generation. These findings have the potential to lead
to highly efficient BWOs with very high output power and excellent
spectral purity.
\end{abstract}


\section{Introduction}

Exceptional points of degeneracy (EPDs) are points in parameter space
of a system at which two or more eigenmodes coalesce in their eigenvalues
(wavenumbers) and eigenvectors (polarization states). Since the characterizing
feature of an exceptional point is the strong degeneracy of at least
two eigenmodes, as implied in \cite{berry2004physics}, we stress
the importance to referring to it as a \textquotedblleft degeneracy\textquotedblright .
Despite most of the published work on EPDs is related to PT symmetry
\cite{bender1998real,heiss1998collectivity,el2007theory,guo2009observation,bittner2012p}
the occurrence of an EPD actually does not require a system to satisfy
PT symmetry. Indeed, EPDs have been recently found also in single
resonators by just adopting time variation of one of its components
\cite{HReza}. EPDs are also found in uniform waveguides at their
cutoff frequencies \cite{mealy2019degeneracy} and in periodic waveguides
at the regular band edge (RBE) and at the degenerate band edge (DBE).
However, these RBE and DBE are EPDs realized in lossless structures
\cite{F1,F2,F3,othman2017experimental,nada2017theory,sloan2018theory}.
Here, we investigate an EPD that requires both distributed power extraction
and gain being simultaneously present in a waveguide called here as
``slow wave structure'' (SWS) since its mode is used to interact
with an electron beam. Note that the passivity of the waveguide here
is not dominated by dissipative losses but rather from power that
is extracted in a continuous fashion from the SWS. Therefore, modes
that propagate in such a waveguide experience exponential decay while
they propagate, as if the waveguide was lossy. A particular and well
studied case of simultaneous existence of symmetric gain and loss
is based on Parity-time (PT)-symmetry, which is a special condition
that leads to the occurrence of an EPD \cite{bender1998real,heiss1998collectivity,el2007theory,guo2009observation,bittner2012p},
that however requires a spatial symmetry in gain and loss. The EPD
considered here is far from that condition, involving two completely
different media that support waves, a plasma and a waveguide for electromagnetic
waves, but still require their interaction and the simultaneous presence
of gain and ``loss''. We stress that in this paper the term ``loss''
is not associated to the damping of energy, but it is rather referred
to a waveguide perspective where energy exits the waveguide in a distributed
fashion, a mechanism referred to as distributed power extraction (e.g.,
distributed radiation) from the interaction zone, for example by realizing
a distributed long slot along the SWS or a set of periodically spaced
holes as in Fig. 1.

In this paper the linear electron beam is modeled using the description
presented in\cite{bohm1949theory}. Then we use the well established
Pierce model \cite{pierce1951waves} to account for the interaction
of the electromagnetic (EM) wave in the SWS and the electron beam,
assuming small signal modulation of the beam. The Pierce model consists
of an equivalent transmission line (TL) governed by telegrapher's
equations, coupled to a plasma-like medium (the linear electron beam)
governed by the equations that describe the electron beam dynamics.
The distributed radiation coming out of the SWS is conventionally
represented by a distributed equivalent ``radiation resistance''
in the TL, following the well established terminology used in Antenna
Theory \cite{hansen2009phased,stutzman2012antenna,wang2006broadband}.
Recently, two coupled transmission lines with balanced gain and loss
(balanced refers to the combination that generates an EPD) were shown
to support EPDs without the need of PT-symmetry \cite{othman2017theory,abdelshafy2018exceptional}.
Even farther from the usual PT-symmetry condition, in this paper the
EPD is generated by a TL supporting backward EM wave propagation that
interacts with a linear electron beam which is a plasma medium that
supports two non-reciprocal waves. We show how this new EPD condition
can be used in high power electron beam devices. Backward-wave oscillators
(BWOs) are widely used as high power sources in radars, satellite
communications, and various other applications. A BWO is composed
of a SWS that guides EM backward waves, where their phase and group
velocities have opposite directions. The interaction between the electron
beam and a backward EM mode constitutes a distributed feedback mechanism
that makes the whole system unstable at certain frequencies \cite{johnson1955backward,tsimring2006electron}.
One of the most challenging issues in BWOs is the limitation in power
generation level, i.e., the extracted power from the electron beam
relative to its total power. Indeed conventional BWOs exhibit small
starting beam current (to induce sustained oscillations) and limited
power efficiency, without reaching very high output power levels.
The extracted power in a conventional BWO is taken from one of the
SWS waveguide ends. In this paper we also propose to use the new EPD
to conceive high power sources. The proposed ``degenerate synchronization
regime'' of operation in a BWO is based on the EPD generated by the
simultaneous presence of gain (coming from the electron beam) and
continuous distributed power extraction from the EM guided wave (rather
than power extraction from the waveguide end). Therefore we stress
that the term ``loss'' does not refer to material loss but rather
to a distributed power extraction mechanism (Fig. 1) that in antennas
terminology is referred to as ``radiation loss'', while the gain
is provided by the electron beam interacting with the SWS. The distributed
power extraction from the interaction zone occurs either via a distributed
set of radiating slots along the SWS (Fig. 1), or alternatively by
collecting the distributed extracted power via an adjacent coupled
waveguide.

Recently, SWSs exhibiting a degenerate band edge (DBE), which is an
EPD of order four in a \textit{lossless} periodic waveguide, were
proposed to enhance the performance of high power devices \cite{othman2016low,F8,abdelshafy2018electron,othman2016theory}.
It is important to point out that the DBE discussed in these previously
mentioned works were obtained in the \textquotedbl cold\textquotedbl{}
SWS by exploiting periodicity, and the benefit of such DBE would gracefully
vanish while increasing the beam power. In this paper, instead, we
propose a new interaction regime where the EPD is maintained in the
\textquotedbl hot\textquotedbl{} SWS, i.e., in presence of the interacting
electron beam, that is a very large source of distributed gain. The
degenerate synchronization between the charge wave induced on the
electron beam and the EM slow wave is occurring at the EPD, making
this degenerate synchronization regime a very special condition not
explored previously.

\section{System model}

\begin{figure}
\centering \subfigure[]{\includegraphics[width=0.26\textwidth]{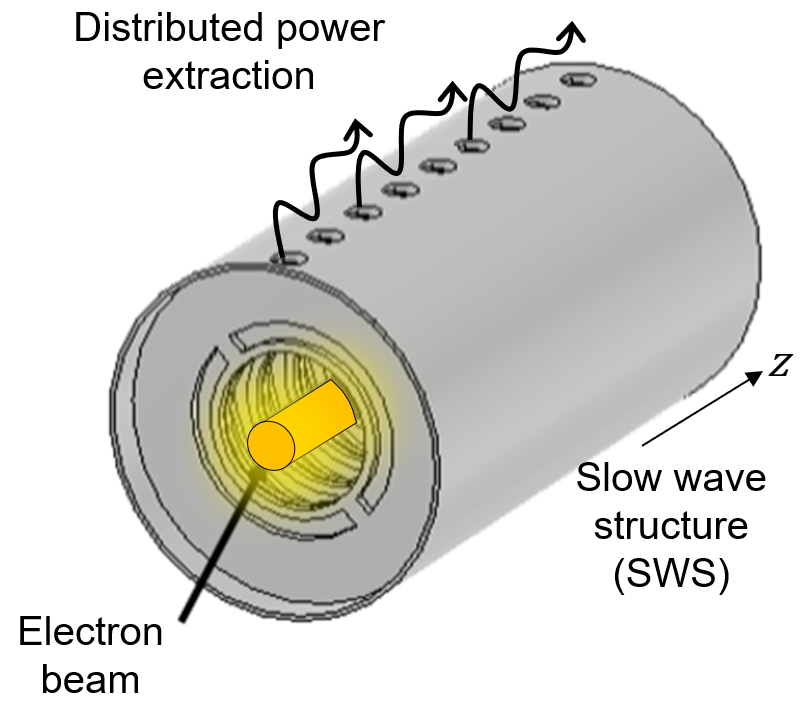}}
\label{Fig:SWS_G} \subfigure[]{\includegraphics[width=0.22\textwidth]{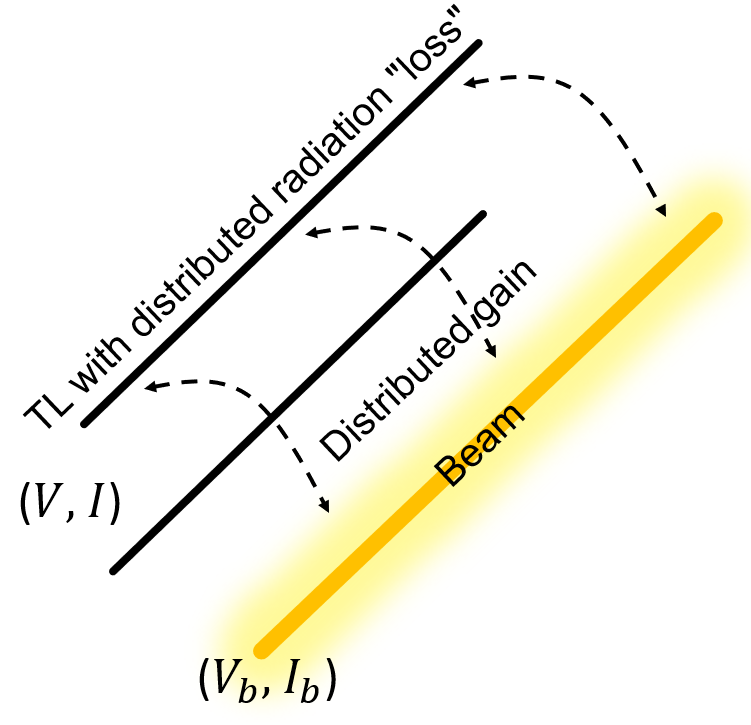}}
\label{Fig:SWS_TL} \caption{(a) An example of SWS with distributed radiating slots that extract
power from the guided modes interacting with the electron beam. (b)
Pierce-based equivalent transmission line model of the SWS with distributed
``loss'' (i.e., radiation) coupled to the charge wave modulating
the electron beam. From the transmission line point of view, the interaction
with the beam is seen as distributed gain. }
\end{figure}
Consider the SWS shown in Fig. 1(a) supporting a Floquet-Bloch mode
whose slow wave spatial harmonic interacts with an electron beam,
and radiates via a distributed set of slots (radiation occurs via
a fast wave spatial harmonic, usually the so called ``-1'' harmonic).
The configuration in Fig. 1(a) is given as a possible realization,
though another one may consist in collecting the power exiting the
interaction zone in a distributed fashion by an adjacent waveguide.
We assume that the interaction between the SWS mode and the electron
beam is modeled via the Pierce small-signal theory of traveling-wave
tubes \cite{bohm1949theory,pierce1947theory,ramo1939space,kompfner1947traveling,pierce1951waves,chu1948field,tamma2014extension}
that describes the evolution of the EM fields and electron beam dynamics
assuming small signal modulation in the beam's electron velocity and
charge density. The beam electrons have average velocity and linear
charge density $u_{0}$ and $\rho_{0}$, respectively. The electron
beam has an average (dc) current $I_{0}=-\rho_{0}u_{0}$ and a dc
(time-average) equivalent kinetic dc voltage $V_{0}=\tfrac{1}{2}u_{0}^{2}/\eta$
\cite{tsimring2006electron}, \cite{gilmour2011klystrons}, and $\eta=e/m=1.758820\text{\ensuremath{\times}}10^{11}$
C/Kg is the charge-to-mass ratio of the electron with charge equal
to $-e$ and $m$ is its rest mass and $\rho_{0}$ is negative. The
small signal modulation in electron beam velocity and charge density
$u_{b}$ and $\rho_{b},$ respectively, form the so called ``charge
wave''. The linearized basic equations governing charge motion and
continuity are written in their simplest form as \cite{pierce1951waves}

\begin{equation}
\begin{array}{c}
\partial_{t}u_{b}+u_{0}\partial_{z}u_{b}=-\eta e_{z},\\
\partial_{t}\rho_{b}=-\rho_{0}\partial_{z}u_{b}-u_{0}\partial_{z}\rho_{b},
\end{array}
\end{equation}
where $e_{z}$ is the electric field component in the $z$-direction
of the EM mode in the SWS interacting with the electron beam, and
the operator $\partial_{\sigma}$ indicates differentiation with respect
to the variable $\sigma$. For convenience, we define the equivalent
kinetic beam voltage and current as $v_{b}=u_{b}u_{0}/\eta$ and $i_{b}=u_{b}\rho_{0}+u_{0}\rho_{b}.$
Assuming an $e^{i\omega t}$ time dependence for monochromatic fields,
the two equations in (1) are written in terms of the beam equivalent
voltage and current in phasor domain as

\begin{equation}
\begin{array}{c}
\partial_{z}V_{b}=-i\beta_{0}V_{b}-E_{z},\\
\partial_{z}I_{b}=-igV_{b}-i\beta_{0}I_{b},
\end{array}
\end{equation}
where $\beta_{0}=\omega/u_{0}$ is the beam's equivalent propagation
constant and $g=\tfrac{1}{2}I_{0}\beta_{0}/V_{0}$. Small fonts are used for the time-domain representation while capital ones are
used for the phasor-domain representation. The details of the model
used to describe the electron beam and its fundamental equations that
describe the beam's dynamics in space and time are given in Appendix
A.

The EM mode propagating in the SWS is described by the equivalent
transmission line in Fig. 1(b), with distributed per-unit-length series
impedance $Z$ and shunt admittance $Y$, and equivalent voltage $V(z)$
and current $I(z)$ phasors that satisfy the telegrapher's equations

\begin{equation}
\begin{array}{c}
\partial_{z}V=-ZI,\\
\partial_{z}I=-YV+I_{s}.
\end{array}\label{eq:TL_Model_Eq}
\end{equation}
The term $I_{s}$ in (\ref{eq:TL_Model_Eq}) accounts for the electron
stream flowing in the SWS that loads the TL as a shunt displacement
current according to \cite{pierce1951waves,ramo1939space,tamma2014extension}
and whose expression is given by $I_{s}=-\partial_{z}I_{b}$. For
the non-interactive EM system (i.e., when $I_{s}$=0), the propagation
constant and the characteristic impedance of the electromagnetic mode
are given by $\beta_{p}=\sqrt{-ZY}$ and $Z_{c}=\sqrt{Z/Y}$, respectively.
The two root solutions represent waves that propagate in opposite
directions, i.e., both $\beta_{p}$ and $-\beta_{p}$ are valid solutions
because of reciprocity. In this paper we consider that the SWS supports
either a ``forward'' or a ``backward'' mode. The modal propagation
constant is a complex number, $\beta_{p}=\beta_{pr}+i\beta_{pi},$where
$\beta_{r}\beta_{i}<0$ for forward modes and $\beta_{r}\beta_{i}>0$
for backward modes. For example, when the the TL is supporting forward
wave propagation both roots in $\beta_{p}$ and $Z_{c}$ must be taken
as the principal square roots, so that their real part is positive.
In this case $\beta_{p}$ turns out to be in in the fourth complex
quadrant \cite{ziolkowski2001wave}. However, for TL supporting backward
wave propagation the root of $\beta_{p}=\sqrt{-ZY}$ is taken as the
principle square root (resulting in a positive real part) while the
root of $Z_{c}=\sqrt{Z/Y}$ is taken as the secondary square root
(resulting in a negative real part). In this case $\beta_{p}$ turns
out to be in in the first complex quadrant\cite{ziolkowski2001wave}.

Based on the Pierce's model \cite{pierce1951waves}, the EM wave couples
to the electron beam with its longitudinal electric field given by
$E_{z}=-\partial_{z}V$. For convenience, we define a state vector
$\mathbf{\boldsymbol{\Psi}}(z)=[\begin{array}{cccc}
V(z), & I(z), & V_{b}(z), & I_{b}(z)\end{array}]^{T}$ that describes the system evolution with coordinate $z$. Thus, the
interacting EM mode and electron-beam charge wave are described as
\cite{tamma2014extension}

\begin{equation}
\partial_{z}\mathbf{\boldsymbol{\Psi}}(z)=-i\mathbf{\underline{\mathbf{M}}}\mathbf{\boldsymbol{\Psi}}(z),\label{eq:Sys_Equation_All}
\end{equation}
where $\mathbf{\underline{\mathbf{M}}}$ is the $4\times4$ system
matrix
\begin{equation}
\underline{\mathbf{M}}=\left[\begin{array}{cccc}
0 & -iZ & 0 & 0\\
-iY & 0 & -g & -\beta_{0}\\
0 & -iZ & \beta_{0} & 0\\
0 & 0 & g & \beta_{0}
\end{array}\right].\label{eq:Sys_Matrixx}
\end{equation}
This description in terms of a multidimensional first order differential
equation in (\ref{eq:Sys_Equation_All}) is ideal for exploring the
occurrence of EPDs in the system since an EPD is a degeneracy associated
to two or more coalescing eigenmodes, hence it occurs when the system
matrix $\underline{\mathbf{M}}$ is similar to a matrix that contains
a non-trivial Jordan block. In general there are four independent
eigenmodes, and each eigenmode is described by an eigenvector $\mathbf{\boldsymbol{\Psi}}(z)$,
hence it includes both the EM and the charge wave. At the EPD investigated
in this paper two of these four eigenvectors coalesce.

\section{Second order EPD in an interacting electromagnetic wave and an electron
beam's charge wave}

Assuming a state vector \textit{z}-dependence of the form $\boldsymbol{\Psi}(z)\propto e^{-ikz}$,
where $k$ is the wavenumber of a mode in the interacting system,
i.e., in the hot SWS, the eigenmodes are obtained by solving the eigenvalue
problem $k\mathbf{\boldsymbol{\Psi}}(z)=\underline{\mathbf{M}}\mathbf{\boldsymbol{\Psi}}(z)$,
and the modal dispersion relation is given by

\begin{equation}
\begin{array}{c}
D(\omega,k)=\mathrm{det}\left(\mathbf{\underline{\mathbf{M}}}-k\mathbf{\underline{\mathbf{I}}}\right)\ \ \ \ \ \ \ \ \ \ \ \ \ \ \ \ \ \ \ \ \ \ \ \ \ \ \ \ \ \ \ \ \ \ \ \ \ \ \ \ \ \ \ \\
=k^{4}-2\beta_{0}k^{3}+\left(\beta_{0}^{2}+ZY-iZg\right)k^{2}\\
\ \ \ \ \ \ \ \ \ \ \ \ \ \ \ \ \ \ \ \ \ \ \ \ \ \ \ \ \ \ \ \ \ \ -2\beta_{0}ZYk+\beta_{0}^{2}ZY=0.
\end{array}\label{eq:Disp_1}
\end{equation}

The solution of this equation leads to four modal complex wavenumbers
that describe the four modes in the EM-electron beam interactive system.
A second order EPD occurs when two of these eigenmodes coalesce in
their eigenvalues and eigenvectors, which means that the matrix $\underline{\mathbf{M}}$
is similar to a matrix that contains a Jordan block of order two \cite{othman2017theory,abdelshafy2018exceptional}.
Following the theory in \cite{hanson2018exceptional}, the algebraic
formulation that is often used to determine EPDs (and summarized in
Appendix B) is equivalent to a bifurcation theory that is here applied.
Indeed the EPD radian frequency and wavenumber are simply obtained
by setting $D(\omega_{e},k_{e})=0$ and $\partial_{k}D(\omega_{e},k)\big|_{k_{e}}=0$
as in \cite{hanson2018exceptional}, where the EPD is designated with
the subscript $e$. Based on the details provided in Appendix B these
two conditions show that the EPD occurs when the transmission line
per-unit-length series impedance and shunt admittance are $Z=Z_{e}$
and $Y=Y_{e}$, where 
\begin{equation}
\begin{array}{ccc}
Z_{e}=\dfrac{i\beta_{0e}^{2}\delta_{e}^{3}}{g_{e}}, &  & Y_{e}=\dfrac{ig_{e}(\delta_{e}+1)^{3}}{\delta_{e}^{3}},\end{array}\label{eq:EPD_Cod}
\end{equation}
and $\delta_{e}=(k_{e}-\beta_{0e})/\beta_{0e}$ represents the relative
deviation of degenerate modal wavenumber $k_{e}$ (of the interactive
system) from the beam equivalent propagation constant $\beta_{0e}=\omega_{e}/u{}_{0}$. 

Using (\ref{eq:EPD_Cod}), we constrain $\delta_{e}$ to provide $Z_{e}$
and $Y_{e}$ with positive real part, that means we assume that the
TL is passive because of SWS losses and especially because of the
distributed power extraction mechanism (see Appendix B). Figure 2
shows the two sectoral regions of $\delta_{e}$ that allow EPDs for
transmission line that support either forward or backward propagation.
These two regions satisfy the passivity condition of the TL, $\mathrm{Re}(Z)\geq0$
and $\mathrm{Re}(Y)\geq0$, where the real part of $Z$ and $Y$ represents
mainly distributed power extraction (i.e., radiation losses using
a terminology in the antennas community). It is important to point
out that dark blue regions in Fig. 2 correspond to complex values
of $\delta_{e}$ where EPD is obtained for TL that has also gain (independently
of the electron beam). As explained in Appendix B, the two light blue
and red narrow sectoral regions in Fig. 2 also correspond to an electron
beam that delivers energy to the TL. In the rest of this paper we
focus on the light blue region that represents complex values of $\delta_{e}$
associated to EPDs resulting from the interaction of a backward propagating
electromagnetic wave and the charge wave modulating the electron beam.
We stress that distributed radiated power is represented by series
and/or parallel ``losses'' in the passive TL.

We now simplify the two equations given in (\ref{eq:EPD_Cod}) to
get rid of $\delta_{e}$ as follows: from the first equation in (\ref{eq:EPD_Cod})
we find that $\delta_{e}=\sqrt[3]{-ig_{e}Z_{e}/\beta_{0e}^{2}}$,
where the proper choice of root, based on chart in Fig. 2, is the
one that guarantees the passivity of the TL and that there is power
delivered from the beam to the TL (see Appendix B). For example, the
root of $\sqrt[3]{-ig_{e}Z_{e}/\beta_{0e}^{2}}$ should lie in the
light blue region in the chart to have an EPD with passive TL that
support backward wave. Roots of $\sqrt[3]{-ig_{e}Z_{e}/\beta_{0e}^{2}}$
that lie in the dark blue region are ignored because they would result
in an EPD that requires an active TL, while here we consider only
passive TLs because of power extraction. The EPD conditions in (\ref{eq:EPD_Cod})
are simplified by substituting the previous expression of $\delta_{e}$
in the second equation of (\ref{eq:EPD_Cod}), leading to an interesting
equation that constrains the TL parameters and the electron beam parameter
$g_{e}$:

\begin{equation}
-Z_{e}Y_{e}/\beta_{0e}^{2}=\left(\sqrt[3]{-ig_{e}Z_{e}/\beta_{0e}^{2}}+1\right)^{3}.\label{eq:EPD_Condition_Free_Delta}
\end{equation}

The above equation constrains all the system parameters to have an
EPD, hence it says that for an EPD to occur, a specific choice of
the electron beam current $I_{0}=I_{0e}$ and angular frequency $\omega=\omega_{e}$
must be selected. In other words it says that for a given electron
beam described by $\beta_{0e}$, and $g_{e}$, the cold TL parameters
$Z_{e}$ and $Y_{e}$ must be chosen accordingly. The condition in
(\ref{eq:EPD_Condition_Free_Delta} can also be rewritten in term
of the propagation constant $\beta_{pe}$ and the characteristic impedance
$Z_{ce}$ of the cold TL as $(\beta_{pe}/\beta_{0e})^{2}=\left(\sqrt[3]{g_{e}Z_{ce}\beta_{pe}/\beta_{0e}^{2}}+1\right)^{3}.$

In general, the interaction between the charge wave and the EM wave
occurs when they are synchronized, i.e., by matching the EM wave phase
velocity $\omega/\beta_{p}$ to the average velocity of the electrons
$u_{0}=\omega/\beta_{0}$, a condition that is specifically called
``synchronization''. Note that this is just an initial criterion,
because the phase velocity of the modes in the \textit{interactive}
systems are different from $\omega/\beta_{p}$ and $u_{0}$. When
the system parameters are such that equation (\ref{eq:EPD_Condition_Free_Delta})
is satisfied and hence an EPD occurs, there are two modes (in the
interactive system) that have exactly the same phase velocity $\omega/\mathrm{Re}(k_{e})$.
Since this synchronization condition corresponds to an EPD, the two
modes in the interactive system are actually identical also in their
eigenvectors $\boldsymbol{\Psi}$. Note that the spatial \textit{z}-evolution
of the interacting EM and electron charge wave is described by four
modes that are solutions of (\ref{eq:Sys_Equation_All}), three of
which have a positive real part of $k$, as discussed next and in
Appendix B. At the EPD two of these four modes coalesce, i.e., they
become identical, that is why we refer to this condition as ``degenerate
synchronization''. In this paper we explore and enforce this very
special kind of synchronization based on an EPD. In particular we
enforce two of the resulting eigenmodes to fully coalesce in both
their wavenumber and system state variable $\mathbf{\boldsymbol{\Psi}}(z)$.
This condition is a new kind of EPD found in physical systems since
it involves the coupling between two different media of propagation,
and may be very promising for achieving new regimes of operation in
electron beam devices with features not obtainable in conventional
regimes.

The EPD condition obtained in (\ref{eq:EPD_Condition_Free_Delta})
guarantees that the system has two repeated eigenvalues and two coalesced
eigenvectors (see Appendix B)

\begin{equation}
\begin{array}{c}
k_{e}=\sqrt[3]{\beta_{0e}\beta_{pe}^{2}},\\
\\
\mathbf{\Psi}_{e}=[\begin{array}{cccc}
1, & ik_{e}/Z, & 1/\delta_{e}, & g_{e}/(\beta_{0e}\delta_{e}^{2})\end{array}]^{T},
\end{array}
\end{equation}
that form the ``degenerate synchronization''. For the interacting
system of EM and charge wave, the EPD represents a point in parameter
space at which the system matrix $\mathbf{M}$ in (\ref{eq:Sys_Matrixx})
is not diagonalizable and it is indeed similar to a matrix that contains
a $2\times2$ Jordan block. This implies that the solution of (\ref{eq:Sys_Equation_All})
includes an algebraic linear growth factor resulting in unusual wave
propagation characteristics as discussed next.

Solution of (\ref{eq:Disp_1}) leads to four eigenmode wavenumbers
and due to the highly non-reciprocal physical nature of the electron
beam, three have positive real part (i.e., $\mathrm{Re}(k)>0$) and
one has it negative. According to the traveling-wave tube theory in
\cite{pierce1947theoryTWT}, \cite{pierce1951waves}, conventionally
applied to BWOs \cite{johnson1955backward}, these three distinct
wavenumbers with $\mathrm{Re}(k)>0$ participate to the synchronization
mechanism, and the electric field in the SWS is generally represented
as

\begin{equation}
V(z)=V_{1}e^{-ik_{1}z}+V_{2}e^{-ik_{2}z}+V_{3}e^{-ik_{3}z}.
\end{equation}
Remarkably, at the EPD the interactive system has two modes with the
same degenerate wavenumber $k_{e}$, resulting in a guided electric
field with linear growth factor as

\begin{equation}
V(z)=zV_{1}e^{-ik_{e}z}+V_{2}e^{-ik_{e}z}+V_{3}e^{-ik_{3}z}
\end{equation}
which is completely different from any other regime of operation.
A degenerate mode described by $\mathbf{\boldsymbol{\Psi}_{\mathit{e}}}$
is composed also of the charge wave, which implies that also the charge
wave propagation is described as in (11), with two terms having the
same degenerate wavenumber $k_{e}$, and one of them exhibiting the
algebraic linear growth besides the exponential behavior.

\begin{figure}
\centering \includegraphics[width=0.48\textwidth]{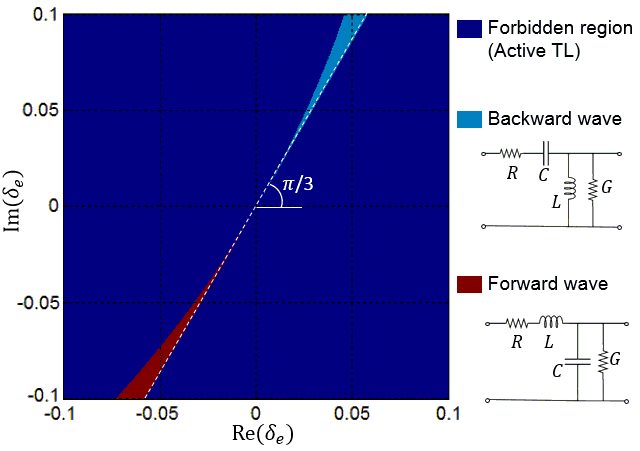}
\label{fig:1a} \caption{The two sectoral regions represent complex values of $\delta_{e}$,
associated to EPDs resulting from the interaction of backward or forward
electromagnetic waves (radiation is represented in TL by series or/and
parallel losses) and the charge wave of electron beam. The value $\delta_{e}=(k_{e}-\beta_{0})/\beta_{0}$
represents the wavenumber deviation of the EPD complex wavenumber
$k_{e}$ (of the interactive system) from the equivalent electron
beam wavenumber $\beta_{0}$, that satisfies (\ref{eq:EPD_Cod}),
assuming SWS realizations based on passive equivalent TLs. There are
two regions of possible realizations (red and light blue), associated
to the two distributed per-unit-length TL circuits on the right panel,
that represent SWSs supporting either forward or backward wave propagation.
In this paper we focus on EPDs obtained from backward electromagnetic
waves interacting with an electron beam, i.e., those leading to $\mathrm{Re}(\delta_{e})>0$
and $\mathrm{Im}(\delta_{e})>0$.}
\end{figure}
In the following we show an example of degenerate synchronization
based on an electron beam with dc voltage of $V_{0}=23$ kV and dc
current of $I_{0}=0.1$ A. We derive the system parameters assuming
that the EPD electron beam current $I_{0}=I_{0e}=0.1$ A (hence the
choice of the TL parameters are chosen accordingly) and assume the
operational frequency at which the EPD occurs is 1 GHz. We require
(as an example) a relative degenerate wavenumber deviation of $\delta_{e}=0.01+i0.017$
that is lying exactly on the white dashed line shown in Fig. 2, with
$\angle\delta_{e}=\pi/3$ (the phase of $\delta_{e}$). The corresponding
wavenumber of two coalescing modes in the interactive system is $k_{e}=\left(1+\delta_{e}\right)\beta_{0e}$
which is $k_{e}=\left(1.01+i0.017\right)\beta_{0e}=70.55+i1.21$ $\mathrm{m}^{-1}$.
By imposing the beam parameters and the chosen $\delta_{e}$ in the
EPD conditions in (\ref{eq:EPD_Cod}) we obtain the distributed per-unit-length
TL impedance and admittance of SWS to be $Z_{e}=-i257.1$ Ohm$\,\mathrm{m}^{-1}$
(i.e., capacitive) and $Y_{e}=1-i19.54$ Siemens$\,\mathrm{m}^{-1}$
(i.e., inductive with losses), representing a backward wave in the
SWS. Losses here are not modeling energy damping but rather energy
extraction (e.g., radiation) from the TL, per unit length, like in
a backward leaky wave antenna \cite{jackson2008leaky}. To obtain
these values, we have chosen the per-unit-length TL parameters for
backward wave propagation as $C=0.62$ pFm, $L=8.14$ pHm, $R=0$
Ohm$\,\mathrm{m}^{-1}$ and $G=1$ Siemens$\,\mathrm{m}^{-1}$. Figure
3(a) shows the dispersion relation of three complex modes in the ``hot''
SWS, i.e., in the interactive EM wave-electron beam system, obtained
using the Pierce model as explained in \cite{tamma2014extension}.
(The fourth mode with $\mathrm{Re}(k_{e})<0$ is not shown since it
does not have a significant role in the synchronism). We show the
real and imaginary parts of wavenumbers of three eigenmodes in the
hot SWS versus normalized frequency (red curves), together with the
dispersion of the beam alone (i.e., the beam ``line'' in green that
actually represents two curves since we are neglecting the effect
of the beam plasma frequency here) and of the backward wave in the
cold SWS (blue curve, with negative slope). It is obvious from the
figure that two eigenmodes of the interactive system coalesce at the
EPD frequency $f_{e}$, forming an EPD of order two. Each of these
two eigenmode has an EM wave and a charge wave counterpart, though
far from the EPD frequency they tend to recover the beam line (green
line) and the EM mode in the cold SWS (blue line).

We recall that the SWS is a periodic structure, and it is the slow
wave harmonic of a mode that interacts with the electron beam charge
wave. Though slow waves do not radiate, a spatial Floquet-Bloch harmonic
of the mode is fast and able to radiate through the slots \cite{jackson2008leaky},
justifying the presence of the conductance $G=1$ Siemens$\,\mathrm{m}^{-1}$,
real part of $Y_{e}=1-i19.54$ Siemens$\,\mathrm{m}^{-1}$, in the
TL model.

\begin{figure}
\centering \subfigure[]{\includegraphics[width=0.23\textwidth]{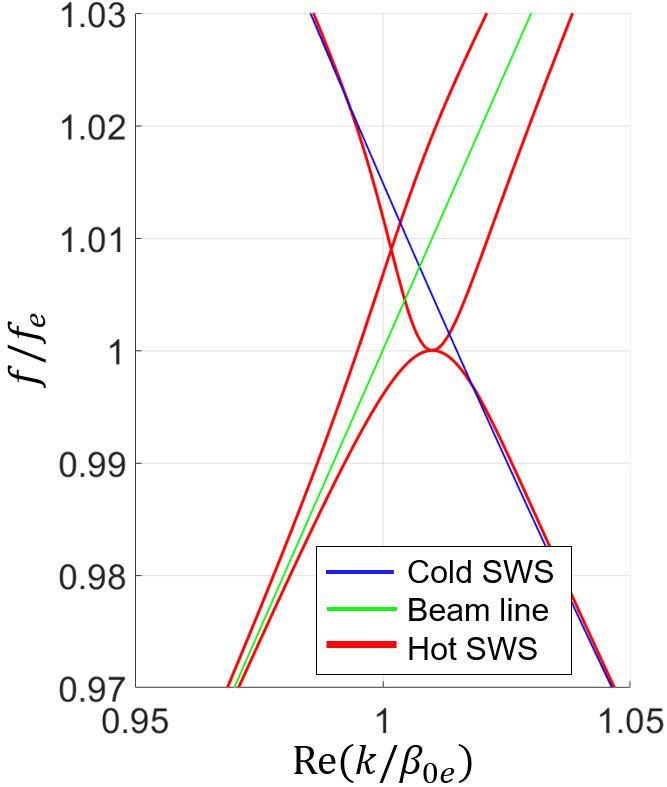}
} \subfigure[]{\includegraphics[width=0.227\textwidth]{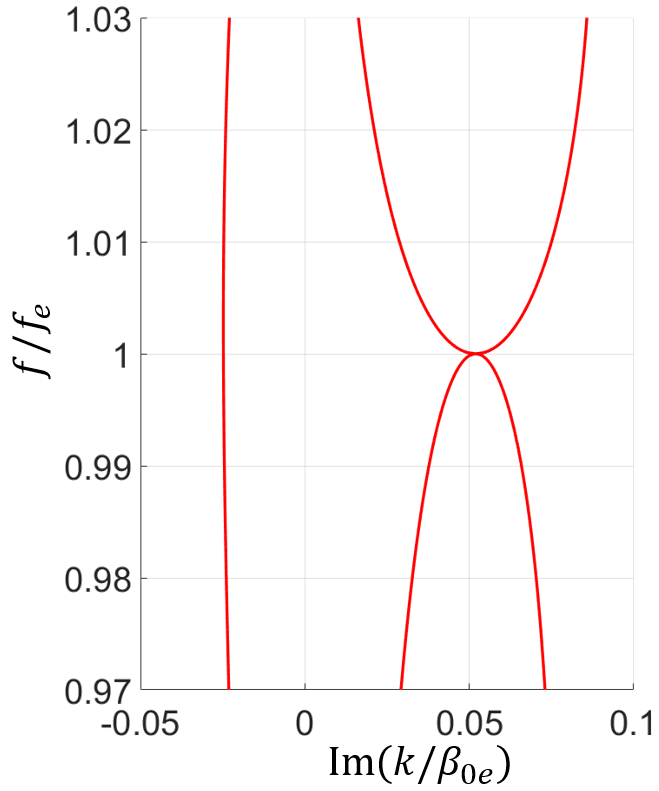}
} \label{Fig:Disp}\caption{Dispersion diagram for three of the four complex modes in the ``hot''
SWS (the modes in the SWS interacting with the electron beam), in
red, showing the existence of a second order EPD, where two modes
coalesce in their wavenumbers and eigenvectors: (a) Real part of the
wavenumber of the modes phase-propagating in the positive $z$ direction
($\mathrm{Re}(k_{e})>0$). The blue line represents the dispersion
of the EM mode in the cold SWS supporting a backward propagation,
whereas the green line is the electron beam's charge wave dispersion,
without accounting for their interaction. (b) Imaginary part of the
wavenumber of the three modes with $\mathrm{Re}(k_{e})>0$ resulting
from the interaction. The EPD wavenumber at $f=f_{e}$ represents
two fully degenerate and synchronous modes with exponential growth
in \textit{z}.}
\end{figure}

\section{EPD-BWO threshold current}

The EPD is here employed to conceive a new regime of operation for
BWOs based on ``degenerate synchronization'' between an EM wave
and the electron beam. To explain this concept we refer to the setup
in Fig. 4(a) for a BWO that has ``balanced gain and radiation-loss''
per unit length, where distributed loss is actually representing distributed
radiation (per unit length) via a shunt conductance $G$ (see Fig.
4). The description is following the steps in \cite{pierce1951waves}
and \cite{johnson1955backward} outlined for BWOs, where the TL is
represented by a distributed circuit model shown in Fig. 4(b) that
supports backward waves. An important difference from a standard BWO
is that here the TL has shunt distributed ``losses'' that actually
model the distributed radiation and whose presence is necessary to
satisfy the EPD condition (\ref{eq:EPD_Condition_Free_Delta}). We
assume to have an unmodulated space charge at the beginning of the
electron beam, i.e., $V_{b}(z=0)=0$ and $I_{b}(z=0)=0$ and we assume
that the SWS waveguide length is $\ell=N\lambda_{e}$, where $\lambda_{e}=2\pi/\beta_{0e}$
is the guided wavelength calculated at the EPD frequency and \textit{N}
is the normalized SWS length. We also assume that the SWS waveguide
is terminated by a load at $z=0$ matched to the characteristic impedance
of the TL (without loss and gain) $R_{o}=\sqrt{L/C}$ and by a short
circuit at $z=\ell$. We follow the same procedure used in \cite{johnson1955backward}
to obtain the \textit{starting oscillation condition} which is based
on imposing infinite voltage gain $A_{v}=V(0)/V(\ell)\to\infty$ .
After simplification, and using the three-wave traveling-wave theory
\cite{pierce1951waves,johnson1955backward}, the voltage gain is written
in terms of the three modes concurring to the synchronization (those
with $\mathrm{Re}(k)>0$ ) as

\begin{equation}
\begin{array}{c}
A_{v}^{-1}e^{i\beta_{0}\ell}=\dfrac{e^{-i\beta_{0}\delta_{1}\ell}\delta_{1}^{2}}{\left(\delta_{1}-\delta_{2}\right)\left(\delta_{1}-\delta_{3}\right)}+\dfrac{e^{-i\beta\delta_{2}\ell}\delta_{2}^{2}}{\left(\delta_{2}-\delta_{3}\right)\left(\delta_{2}-\delta_{1}\right)}\\
\\
\ \ \ \ \ \ \ \ \ \ \ \ \ \ \ \ \ \ \ \ \ \ \ \ \ \ \ \ \ \ \ \ \ \ \ \ \ \ \ +\dfrac{e^{-i\beta_{0}\delta_{3}\ell}\delta_{3}^{2}}{\left(\delta_{3}-\delta_{1}\right)\left(\delta_{3}-\delta_{2}\right)}=0,
\end{array}\label{eq:Gain_EXpress_1}
\end{equation}
where $\delta_{n}=(k_{n}-\beta_{0})/\beta_{0}$ and $k_{1},k_{2}$
and $k_{3}$ are the three wavenumbers of the interactive EM-beam
system with positive real part that are solutions of (\ref{eq:Disp_1}).
In close proximity of the EPD there are two modes coalescing, with
$\delta_{1}=\delta_{a}+\Delta/2$ and $\delta_{2}=\delta_{a}-\Delta/2$
where $\delta_{a}=(\delta_{1}+\delta_{2})/2$ is the average, and
$|\Delta|\ll|\delta_{a}|$ is a very small quantity that vanishes
at the EPD. By observing that $|e^{-i\beta_{0}\delta_{3}\ell}|\ll1$
for very large $\ell$ because Im($k_{3}$)\textless 0, the gain
expression in (\ref{eq:Gain_EXpress_1}) reduces to

\begin{equation}
\begin{array}{c}
A_{ve}^{-1}e^{i\beta_{0}\ell}\approx\dfrac{e^{-i\beta_{0}\delta_{a}\ell}\delta_{a}^{2}}{\left(\delta_{a}-\delta_{3}\right)}\dfrac{\sin\left(\Delta\beta_{0}\ell/2\right)}{\Delta/2}=0\end{array}.\label{eq:Gain_Expr_Final}
\end{equation}

From the above condition, assuming very large $\ell$, the first oscillation
frequency occurs when $\Delta\beta_{0}\ell=2\pi$. This happens when
the constraint on the wavenumbers $k_{1}-k_{2}=2\pi/\ell$ is satisfied.
This shows a very important fact, that for infinitely long structure
$\ell\to\infty$ the starting oscillation condition is $k_{1}=k_{2}=k_{e}$,
which corresponds to the EPD condition. This implies that the EPD
is the exact condition for synchronization between the charge wave
and EM wave, accounting for the interaction in infinitely long SWSs,
that guarantees the generation of oscillations at the EPD frequency
$f_{e}=\omega_{e}/(2\pi)$.

For finite length SWS, oscillations occurs when $k_{1}-k_{2}=2\pi/\ell$
is satisfied, assuming very large $\ell$, and since $|\Delta|\ll|\delta_{a}|$,
we have both $k_{1}$ and $k_{2}$ very close to $k_{e}$, which implies
that the systems is close to the EPD and hence the threshold beam
current $I_{th}$ that starts oscillations is slightly different from
$I_{0e}$. A beam current slightly away from the EPD one causes the
wavenumbers to bifurcate from the degenerate one $k_{e}$, following
the Puiseux series approximation \cite{welters2011explicit} (also
called fractional power expansion) as

\begin{equation}
k_{n}-k_{e}\approx(-1)^{n}\alpha\sqrt{I_{0}-I_{0e}},
\end{equation}
with $n=1,2$ , and $\alpha$ is a constant. This implies that $k_{1}-k_{2}\approx-2\alpha\sqrt{I_{0}-I_{0e}}$,
and by comparing this with the $k_{1}-k_{2}$ difference associated
to the threshold beam current in a finite length SWS we infer that
the threshold beam current $I_{th}$ that makes the EPD-BWO of finite
length \textit{$\ell$} oscillate, asymptotically scales as

\begin{equation}
I_{th}\sim I_{0e}+\left(\dfrac{\pi/\alpha}{\ell}\right)^{2}.\label{eq:Ass_Scaling}
\end{equation}
In a conventional BWO, that has no distributed power extraction and
the power is delivered only to the load $R_{0}$, the threshold current
scaling decrease with the SWS length asymptotically as $I_{th}\sim\zeta/\ell^{3}$
\cite{walker1953starting,johnson1955backward}, where $\zeta$ is
a constant. Instead, the EPD-BWO has a threshold current in (\ref{eq:Ass_Scaling})
always larger than the EPD beam current $I_{0e}$, which represents
the current that keeps the oscillation going and simultaneously balance
the distributed radiated power. Importantly, the EPD beam current
$I_{0e}$ can be engineered to any desired (high) value depending
on how much power one wants to extract from the electron beam.

The above derivation was based on assuming $\ell\to\infty$ (details
are in Appendix D), and a rigorous derivation for any EPD-BWO length
is shown in Appendix C that leads to the determination of the threshold
current and oscillation frequency. This formulation based on the Pierce
model for the beam current threshold is used to compute the results
shown in Fig. 4 for varying the EPD-BWO length and the shunt conductance
\textit{G} that represents the distributed radiation.

Figure 4(c) shows the scaling of the threshold current versus SWS
length for a system with the same parameters used in the previous
example. The scaling shows that for infinitely long SWSs the oscillation
starting current is equal to the EPD beam current $I_{0e}$ which
is consistent with the asymptotic relation in (\ref{eq:Ass_Scaling}).
Figure 4(d) also shows a comparison between the threshold current
of the conventional BWO (that does not have distributed power extraction,
i.e., $G=0$) and that of the EPD-BWO (with distributed power extraction,
represented by $G=1$ Siemens$\,\mathrm{m}^{-1}$) by showing their
current scaling with the SWS length. The threshold beam current $I_{th}$
of the conventional BWO vanishes when the SWS length increases, whereas
the threshold beam current $I_{th}$ of the EPD-BWO tends to the value
of $I_{0e}=0.1$ A. Figure 4(e) shows the threshold beam current $I_{0}=I_{0e}$
(blue curve) that leads to the EPDs as a function of the radiation
``loss'' per unit length \textit{G} . We also show the required
starting beam current (i.e. the threshold) for oscillations (red curve)
for a finite length ``hot'' SWS working at the EPD; we assume the
SWS length normalized to the wavelength at the EPD frequency of $N=70$.
It is important to point out that the radiated power per-unit-length
of the SWS is determined using $p_{rad}(z)=\tfrac{1}{2}G|V(z)|^{2}$,
thus it is linearly proportional to the parameter $G$ , i.e., higher
values of $G$ imply higher radiated power per-unit-length and therefore
higher level of energy extraction from the SWS.

We have shown two very important facts here: first, the threshold
beam current is very close to the EPD beam current for any desired
value of power extracted. This indicates that the EPD condition for
hot SWSs with finite length is the condition that basically guarantees
full synchronization between the EM guided mode and the beam's charge
wave. Secondly, the threshold current increases monotonically when
increasing the required radiated power per unit length which implies
a tight synchronization regime guaranteed for any high power generation.
Therefore in principle the synchronism is maintained for any desired
power output, according to the Pierce model, and this trend is definitely
not observed in standard backward wave oscillators where the load
is at the beginning of the SWS.

\begin{figure}
\centering \subfigure[]{\includegraphics[width=0.35\textwidth]{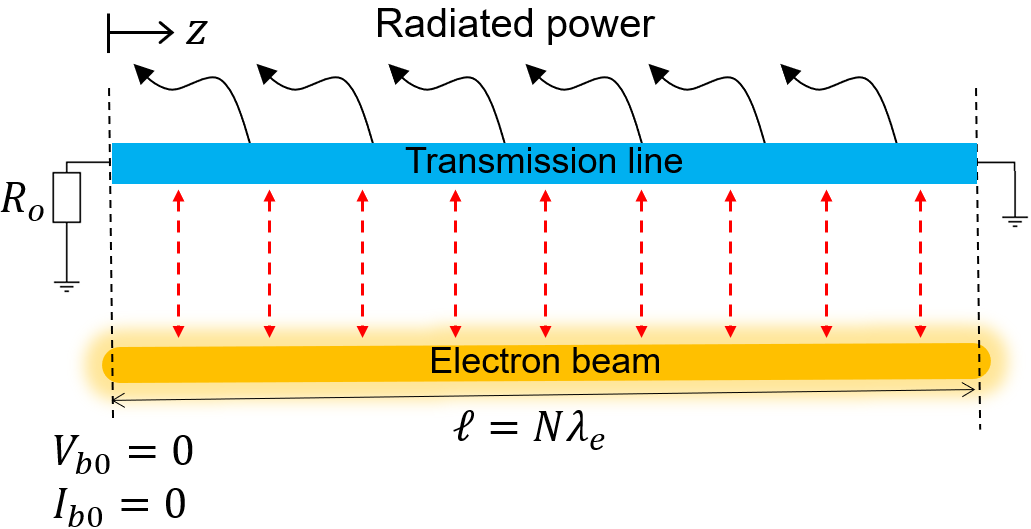}}
\subfigure[]{\includegraphics[width=0.13\textwidth]{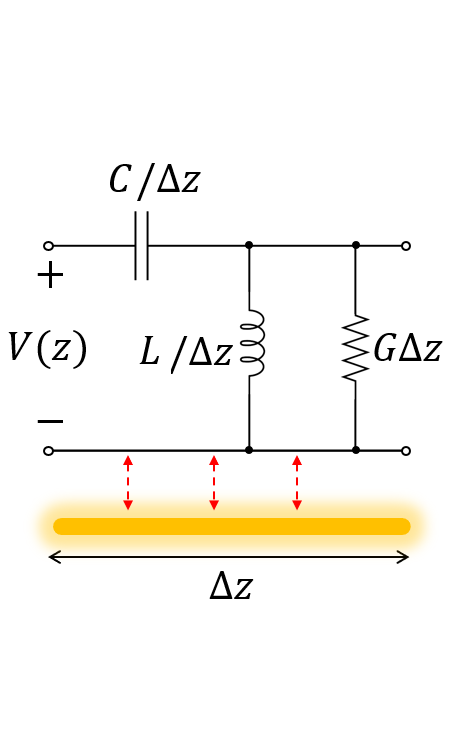}}

\centering \subfigure[]{\includegraphics[width=0.23\textwidth]{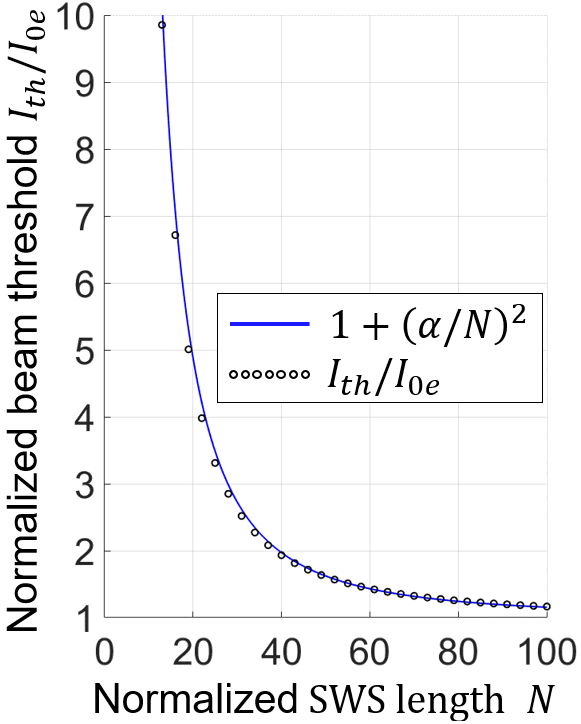}}
\subfigure[]{\includegraphics[width=0.23\textwidth]{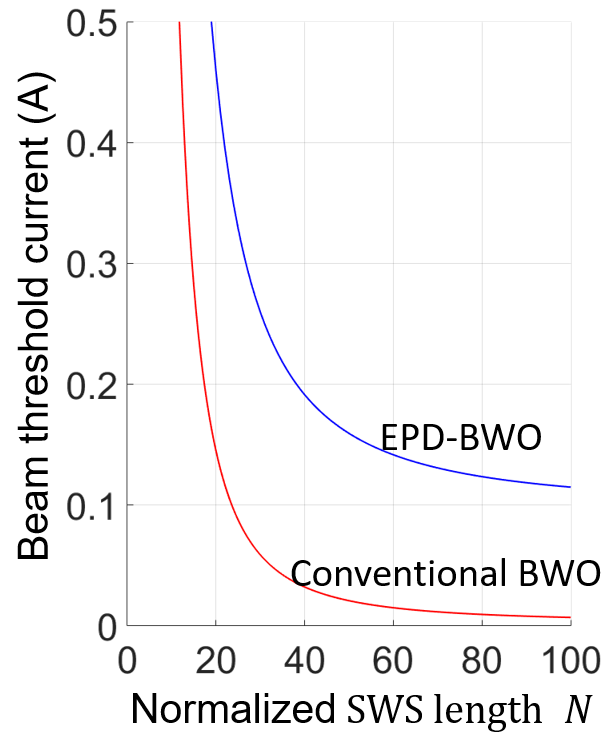}
}
\raggedright{}\centering \subfigure[]{\includegraphics[width=0.26\textwidth]{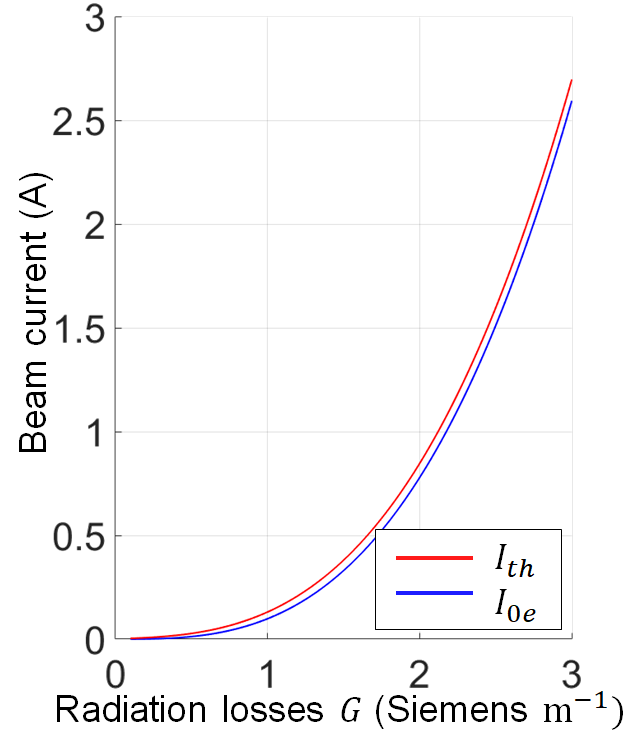}}\caption{(a) Schematic setup for BWOs with ``balanced gain and radiation-loss''.
(b) Equivalent transmission line model of the SWS with distributed
(per-unit-length) series capacitance and shunt inductance for a SWS
that supports backward waves. The distributed shunt conductance \textit{G}
represents distributed power extraction, that is indeed given by $p_{rad}(z)=G|V(z)|^{2}/2$.
(c) Scaling of the oscillation-threshold beam current $I_{th}$ versus
SWS length normalized to the wavelength $N=\ell/\lambda_{e}$, where
$\lambda_{e}=2\pi/\beta_{0e}$ is the guided wavelength calculated
at the EPD frequency. It is obvious that for infinite long SWS where
$N\rightarrow\infty$ we have $I_{th}\rightarrow I_{0e}$ which implies
that the EPD synchronization condition is also the threshold for infinitely
long SWSs. (d) Comparison of the threshold current for an EPD-BWO
and a standard BWO for increasing SWS length. Note that the threshold
of the EPD-BWO does not decreases to zero for longer SWSs, a characteristic
that is fundamentally different from that of a standard BWO. (e) Scaling
of the EPD beam current $I_{0}=I_{0e}$ (blue curve) versus radiation
losses $G$. We also show the scaling of the threshold beam current
(red curve) for a finite-length SWS with $N=70$. Note that the threshold
beam current is very close to the EPD current for any amount of required
distributed extracted power which indicates that in principle the
synchronism is achieved for any output power level.}
\end{figure}

\section{Conclusion}

We have conceptually demonstrated the occurrence of an EPD in an interactive
system made of a linear electron beam and a guided electromagnetic
wave. This EPD condition leads to a new regime of operation for BWOs
where the EPD guarantees a synchronism between a backward wave and
a beam's charge wave through enforcing the coalescence of two modes
in both their wavenumber and state vector, a regime we named ``degenerate
synchronization''. A remarkable aspect of this EPD-BWO regime is
that the ``gain and distributed-power-extraction balance'' condition
leads to a perfect degenerate synchronization between the charge wave
and EM wave for any amount of designed distributed power extraction.
This distributed power extraction can be in the form of distributed
radiation from the interaction zone or of distributed transfer of
power from the hot SWS to an adjacent coupled waveguide. Under this
EPD-BWO regime, in principle it is possible to extract large amounts
of power from the electron beam and therefore the EPD-BWO exhibits
high starting beam current. In theory the starting-oscillation beam
current (i.e, the threshold) could be set to arbitrary large values,
in contrast to what happens in conventional BWOs where the beam's
starting current tends to vanish when the SWS lengths increases. Remarkably,
in the EPD-BWO regime the starting oscillation current is always larger
than the EPD's beam current that, in principle, can be set to large
values by increasing the amount of power extracted per unit length.
Therefore we have shown the fundamental principle that the amount
of power generated under the EPD-BWO regime has no upper limit (the
actual limit would be imposed only by the constraints encountered
in the practical realizability), contrarily to conventional knowledge
of BWOs.

Note that the degenerate synchronization regime discussed in this
paper is very different from the ones discussed in \cite{othman2016low,F8,abdelshafy2018electron,othman2016theory}.
There, it was the ``cold'' SWS that exhibited a degeneracy condition,
like the degenerate band edge (DBE), which is an EPD of order four,
or the stationary inflection point (SIP), which is an EPD of order
three, that were proposed to enhance the performance of high power
devices. Those DBE and SIP EPD conditions were obtained in \textquotedbl cold\textquotedbl{}
SWSs based on periodicity, and indeed the interaction of the EM modes
with the electron beam would perturb those degeneracy conditions,
and even destroy them for increasingly large values of electron beam
currents. Here, instead, we have proposed a new fully synchronous
degenerate regime based on the concept of ``distributed radiation
and gain balance'', where the EPD occurs in the \textquotedbl hot\textquotedbl{}
structure, i.e., in presence of the interacting electron beam with
any amount of current and hence, in principle, for any large amount
of power.

\section{Acknowledgment}

This material is based upon work supported by the Air Force Office
of Scientific Research award number FA9550-18-1-0355. 

\appendices{}

\section{Electron beam model}

We show the fundamental equations that describe the evolution of electron
beam dynamics in space and time. We follow the linearized equations
that describe the space-charge wave on the electron beam presented
in Pierce\cite{pierce1951waves} based on the electron beam model
in \cite{bohm1949theory}. We assume a narrow cylindrical beam of
electrons subject to axial electric field that is constant across
the beam\textquoteright s cross section; we assume purely longitudinal
electron motion as conventionally done in many electron beam devices
thanks to confinement due to applied magnetic field, and negligible
repulsion forces between electrons (hence we neglect space charge's
induced forces) compared to the force induced by the longitudinal
electric field associated to the radio frequency mode in the slow
wave structure (SWS) (this last assumption could be easily removed).
The beam's total linear-charge density and electron speed are represented
as

\begin{equation}
\begin{array}{c}
\rho(z,t)=\rho_{0}+\rho_{b}\\
u(z,t)=u_{0}+u_{b},
\end{array}\label{eq:DC_AC_Charge_Speed}
\end{equation}
where the subscripts '$0$' and '$b$' denote dc (average value) and
ac (alternate current, i.e., modulation), respectively, and $\rho_{0}$
here is negative and $u$ is the electron speed in the $z$-direction.
The basic equations governing the charges' motion and continuity are
written in their simplest form as \cite{pierce1951waves}

\begin{equation}
\begin{array}{c}
\dfrac{du}{dt}=-\eta e_{z}\\
\dfrac{\text{\ensuremath{\partial}}\rho}{\text{\ensuremath{\partial}}t}=-\dfrac{\text{\ensuremath{\partial}}(\rho u)}{\text{\ensuremath{\partial}}z}
\end{array}\label{eq:Basic_Eq_motion_Cont}
\end{equation}
where $\eta=e/m=1.758820\text{\ensuremath{\times}}10^{11}$ C/Kg is
the charge-to-mass ratio of the electron such that the electron charge
is equal to $-e$ and $m$ is its rest mass. Assuming small signal
modulation \cite{ramo1939space,bohm1949theory}, $|u_{b}|\ll u_{0}$
and $|\rho_{b}|\ll|\rho_{0}$\textbar , (\ref{eq:Basic_Eq_motion_Cont})
is linearized to

\begin{equation}
\begin{array}{c}
\dfrac{\partial u_{b}}{\partial t}+u_{0}\dfrac{\text{\ensuremath{\partial}}u_{b}}{\text{\ensuremath{\partial}}z}=-\eta e_{z}\\
\text{\ensuremath{\dfrac{\text{\ensuremath{\partial}}\rho_{b}}{\text{\ensuremath{\partial}}t}}}=-\rho_{0}\dfrac{\partial u_{b}}{\partial z}-u_{0}\dfrac{\partial\rho_{b}}{\partial z}
\end{array}\label{eq:Basic_Eq_motion_Cont_Linear}
\end{equation}

For convenience, we define the equivalent kinetic beam voltage and
current \cite{pierce1951waves} as

\begin{equation}
\begin{array}{c}
v(z,t)=\dfrac{u^{2}}{2\eta}=V_{0}+v_{b},\\
i(z,t)=u\rho=-I_{0}+i_{b}.
\end{array}\label{eq:Voltage_Current_Def}
\end{equation}

Again, assuming small signal modulation \cite{ramo1939space,bohm1949theory},
$|u_{b}|\ll u_{0}$ and $|\rho_{b}|\ll\rho_{0}$, we neglect the terms
$u_{b}^{2}$ and $u_{b}\rho_{b}$ in (\ref{eq:Voltage_Current_Def}).
Thus, the dc and ac beam voltage and current are determined as a function
of the dc and ac beam charge density and speed \cite{tsimring2006electron},
\cite{gilmour2011klystrons}, \cite{pierce1951waves} as

\begin{equation}
\begin{array}{c}
V_{0}=\tfrac{1}{2}u_{0}^{2}/\eta,\\
I_{0}=-\rho_{0}u_{0},\\
v_{b}=u_{b}u_{0}/\eta,\\
i_{b}=u_{b}\rho_{0}+u_{0}\rho_{b}.
\end{array}\label{eq:V_I_Relations}
\end{equation}

By substituting (\ref{eq:V_I_Relations}) into (\ref{eq:Basic_Eq_motion_Cont_Linear}),
the evolution equations of beam equivalent kinetic voltage and current
are written as

\begin{equation}
\begin{array}{c}
\dfrac{\text{\ensuremath{\partial}}v_{b}}{\text{\ensuremath{\partial}}z}=-\dfrac{1}{u_{0}}\dfrac{\text{\ensuremath{\partial}}v_{b}}{\text{\ensuremath{\partial}}t}-e_{z},\\
\dfrac{\text{\ensuremath{\partial}}i_{b}}{\text{\ensuremath{\partial}}z}=\dfrac{\eta\rho_{0}}{u_{0}^{2}}\dfrac{\text{\ensuremath{\partial}}v_{b}}{\text{\ensuremath{\partial}}t}-\dfrac{1}{u_{0}}\dfrac{\text{\ensuremath{\partial}}i_{b}}{\text{\ensuremath{\partial}}t}.
\end{array}\label{eq:1TLbeam-1}
\end{equation}

Assuming time-harmonic signals with time convention $e^{i\omega t}$,
that is omitted in the following for simplicity, equations in (\ref{eq:1TLbeam-1})
are simplified to

\begin{equation}
\begin{array}{c}
\dfrac{\text{\ensuremath{\partial}}V_{b}}{\text{\ensuremath{\partial}}z}=\dfrac{-i\omega}{u_{0}}V_{b}-E_{z},\\
\dfrac{\text{\ensuremath{\partial}}I_{b}}{\text{\ensuremath{\partial}}z}=\dfrac{i\omega\eta\rho_{0}}{u_{0}^{2}}V_{b}-\dfrac{i\omega}{u_{0}}I_{b}.
\end{array}\label{eq:1TLbeam-1-1}
\end{equation}

When the electron beam is not interacting with an electromagnetic
wave, i.e., when $E_{z}=0$, the charge wave has propagation constant
$\beta_{0}=\omega/u_{0}$.

\section{Second order EPD in a system made of an electromagnetic wave interacting
with an electron beam charge's wave}

The solution of the dispersion equation in Eq. (\ref{eq:Disp_1})
has four roots which represent the eigenvalues of the interactive
system composed of guided EM waves and the charge waves that modulate
the electron beam. A necessary condition to have second order EPD
is to have two repeated eigenvalues, which means that the characteristic
equation should have two repeated roots as

\begin{equation}
\begin{array}{c}
D(\omega_{e},k)\propto(k-k_{e})^{2}\end{array}\label{eq:Disp_Req}
\end{equation}
where $\omega_{e}$ and $k_{e}$ are the degenerate angular frequency
and wavenumber, respectively. The relation in (\ref{eq:Disp_Req}),
that guarantees having two coalescing waenumbers, is satisfied when
\cite{hanson2018exceptional}

\begin{equation}
\begin{array}{c}
D(\omega_{e},k_{e})=0,\\
\dfrac{\partial D(\omega_{e},k)}{\partial k}\Bigg|_{k=k_{e}}=0.
\end{array}\label{eq:EPD_Con1}
\end{equation}

Substituting the determinant expresion (\ref{eq:Disp_1}) into (\ref{eq:EPD_Con1}),
the two EPD conditions are 

\begin{equation}
\begin{array}{c}
k_{e}^{4}-2\beta_{0e}k_{e}^{3}+\left(\beta_{0e}^{2}+Z_{e}Y_{e}-ig_{e}Z_{e}\right)k_{e}^{2}\ \ \ \ \ \ \ \ \ \ \ \ \ \ \\
\ \ \ \ \ \ \ \ \ \ \ \ \ \ \ \ \ \ \ \ \ \ \ \ \ -2\beta_{0e}Z_{e}Y_{e}k_{e}+\beta_{0e}^{2}Z_{e}Y_{e}=0,
\end{array}\label{eq:C1}
\end{equation}

\begin{equation}
4k_{e}^{3}-6\beta_{0e}k_{e}^{2}+2\left(\beta_{0e}^{2}+Z_{e}Y_{e}-ig_{e}Z_{e}\right)k_{e}-2\beta_{0e}Z_{e}Y_{e}=0,\label{eq:C2}
\end{equation}
where the EPD is designated with the subscript $e$, i.e., the parameters
with the subscript $e$ are calculated at the EPD frequency; for example,
$\beta_{0e}=\omega_{e}/u_{0}.$ Only at a specific frequency (the
EPD frequency) it is possible to find two identical eigenvalues. Therefore
the above equation provides both the EPD radian frequency $\omega_{e}$
and wavenumber $k_{e}$.

The combination of the TL distributed series impedance $Z=Z_{e}$
and shunt admittance $Y=Y_{e}$ that provide the EPD are determined
after making some mathematical manipulations in the two conditions
in (\ref{eq:C1}) and (\ref{eq:C2}). First, we use (\ref{eq:C1})
to get $Y_{e}$ in term of $Z_{e}$ and other system parameters as

\begin{equation}
Y_{e}=\dfrac{ig_{e}k_{e}^{2}}{\left(k_{e}-\beta_{0e}\right)^{2}}-\dfrac{k_{e}^{2}}{Z_{e}},\label{eq:Ze_Ye_Rel}
\end{equation}
then we substitute this relation in (\ref{eq:C2}) and solve for $Z_{e}$
which is found to be

\begin{equation}
Z_{e}=\dfrac{i\left(k_{e}-\beta_{0e}\right)^{3}}{\beta_{0e}g_{e}}.
\end{equation}
We finally substitute $Z_{e}$ in (\ref{eq:Ze_Ye_Rel}) to obtain
$Y_{e}$

\begin{equation}
Y_{e}=\dfrac{ig_{e}k_{e}^{3}}{\left(k_{e}-\beta_{0e}\right)^{3}}.
\end{equation}
These two latter expressions are then rewritten as in (\ref{eq:EPD_Cod}).
Assuming that the EPD conditions in (28) and (29) are satisfied, the
degenerate wave number $k_{e}$ is determined by the product of (28)
and (29) $Z_{e}Y_{e}=-k_{e}^{3}/\beta_{0e}$ and by recalling that
$Z_{e}Y_{e}=-\beta_{pe}^{2}$ , leading to $k_{e}=\sqrt[3]{\beta_{0e}\beta_{pe}^{2}}$,
which is the result in (9).

The eigenvectors $\mathbf{\mathbf{\Psi}}_{n}$ of the system are determined
by solving

\begin{equation}
\left(\mathbf{\underline{\mathbf{M}}}-k_{n}\mathbf{I}\right)\mathbf{\mathbf{\Psi}_{\mathrm{\mathit{n}}}}=\mathrm{\mathbf{0},}\label{eq:eigenvector_Eq}
\end{equation}
where $k_{n}$ with $n=1,2,3,4$ are the modal wavenumbers, and they
are determined from (\ref{eq:Disp_1}). By solving (\ref{eq:eigenvector_Eq}),
the eigenvectors are written in the form (each element of the eigenvector
carries an implicit unit of Volt besides the units of the explicit
parameters)

\begin{equation}
\mathbf{\mathbf{\Psi}_{\mathrm{\mathit{n}}}}=[\begin{array}{cccc}
1, & ik_{n}/Z, & (1+\delta_{n})/\delta_{n}, & g(1+\delta_{n})/(\beta_{0}\delta_{n}^{2})\end{array}]^{T},\label{eq:eigenvector_1}
\end{equation}
where $\delta_{n}=(k_{n}-\beta_{0})/\beta_{0}$. As shown in \cite{pierce1951waves},
three of these four modes are strongly affected by the synchronization
of the electron beam and the EM mode with positive phase velocity.
Assuming that these three wavenumbers of the beam-EM mode interactive
system are a slight perturbation of the unperturbed beam's propagation
constant, i.e, $k_{n}\approx\beta_{0}$ with, $n=1,2,3$ the eigenvector
expression in (\ref{eq:eigenvector_1}) is approximated

\begin{equation}
\mathbf{\mathbf{\Psi}_{\mathrm{\mathit{n}}}}\approx[\begin{array}{cccc}
1, & ik_{n}/Z, & 1/\delta_{n}, & g/(\beta_{0}\delta_{n}^{2})\end{array}]^{T}.\label{eq:eigenvector2}
\end{equation}

It is worth mentioning that the eigenvector expression in (\ref{eq:eigenvector_1})
is valid for any of the four modes of the interacting system. Whereas
the expression in (\ref{eq:eigenvector2}) is only valid for the three
modes with positive $\mathrm{Re}(k)$, namely, for the three synchronous
modes resulting from the interaction of the SWS EM mode with the electron
beam. Those three synchronous modes are such that $k_{n}\approx\beta_{0}$
based on synchronization. Two of these eigenvectors coalesce at the
EPD, forming the degenerate syncrhosim between the two modes $\mathbf{\mathbf{\Psi}}_{1}=\mathbf{\mathbf{\Psi}}_{2}=\mathbf{\mathbf{\Psi}}_{e}$,
where

\begin{equation}
\mathbf{\Psi}_{e}\approx[\begin{array}{cccc}
1, & ik_{e}/Z, & 1/\delta_{e}, & g_{e}/(\beta_{0e}\delta_{e}^{2})\end{array}]^{T}.
\end{equation}

In summary, the two conditions in (\ref{eq:EPD_Cod}) represent constraints
on the TL parameters, calculated at EPD frequency, that provide the
second order EPD, where two eigenmodes of the interacting system have
identical eigenvalues $k_{1}=k_{2}=k_{e}$ and eigenvectors $\mathbf{\mathbf{\Psi}}_{1}=\mathbf{\mathbf{\Psi}}_{2}=\mathbf{\mathbf{\Psi}}_{e}$.
These two eigenmodes form the degenerate synchronization.

From the transmission line point of view the electron beam is modeled
as parallel per-unit-length admittance 

\begin{equation}
Y_{b}=\dfrac{-igk^{2}}{(k-\beta_{0})^{2}},
\end{equation}
that loads the line \cite{tamma2014extension}. For proper operation
of this new degenerate BWO regime the TL should outcouple power in
a distributed fashion and the electron beam should supply power to
balance the power extraction in order to have a sustainable oscillation.
This means that both the TL parameters $Z$ and $Y$ should be passive,
i.e., $\mathrm{Re}(Z)>0$ and $\mathrm{Re}(Y)>0,$ and the beam equivalent
admittance should be active, i.e., $\mathrm{Re}(Y_{b})<0$. These
constraints are plotted in Fig. 5 as a function of the complex wavenumber
deviation $\delta_{e}=(k_{e}-\beta_{0e})/\beta_{0e}$. The intersection
of these three conditions on $\mathrm{Re}(Z)>0$ and $\mathrm{Re}(Y)>,$
and $\mathrm{Re}(Y_{b})<0$ lead to the realizability diagram in Fig.
5(d) which is the same as in Fig. 2.

\begin{figure}
\centering \subfigure[]{\includegraphics[width=0.22\textwidth]{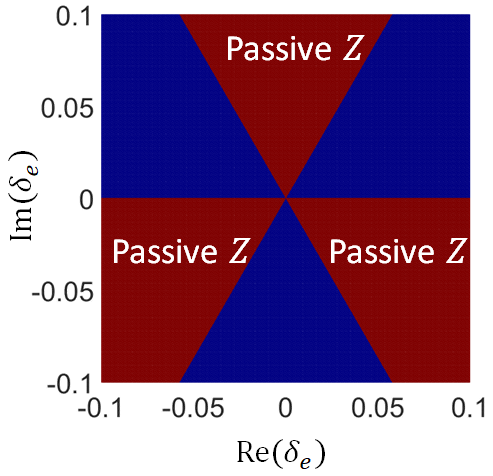}}
\subfigure[]{\includegraphics[width=0.22\textwidth]{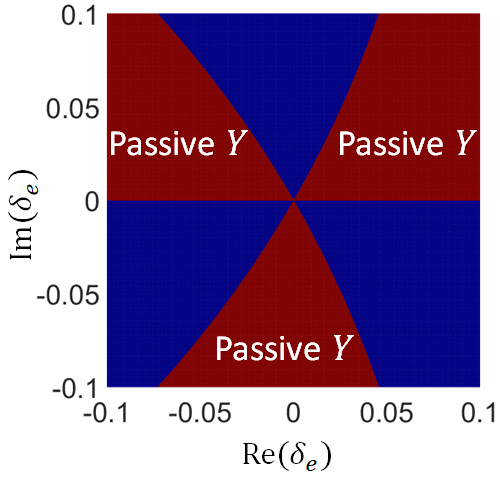}} 

\centering \subfigure[]{\includegraphics[width=0.22\textwidth]{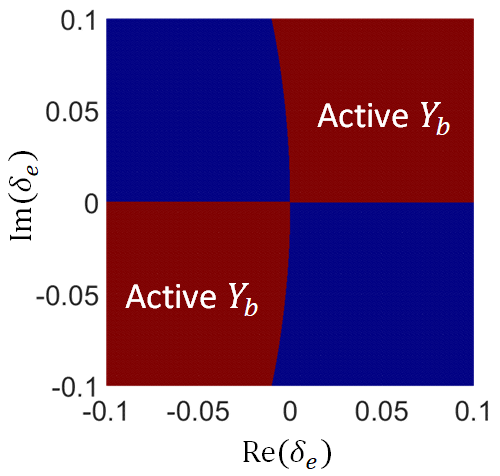}}
\subfigure[]{\includegraphics[width=0.22\textwidth]{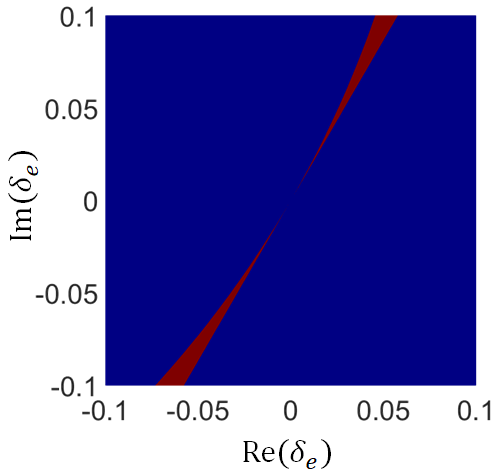}
} \caption{Regions of complex values $\delta_{e}$ that result in having passive
TL, with $\mathrm{Re}(Z)>0$ and $\mathrm{Re}(Y)>0$ when EPD conditions
are satisfied, i.e., when$Z=Z_{e}$ and $Y=Y_{e}$: (a) passive series
impedance $\mathrm{Re}(Z_{e})=\mathrm{Re}(i\beta_{0e}^{2}\delta_{e}^{3}/g_{e})>0$
and (b) passive parallel admittance $\mathrm{Re}(Y_{e})=\mathrm{Re}(ig_{e}(\delta_{e}+1)^{3}/\delta_{e}^{3})>0$.
(c) Regions of complex values $\delta_{e}$ where the electron beam
delivers energy to the TL, i.e., when the electron beam provides gain
from TL perspective, which means that $\mathrm{Re}(Y_{be})<0$, i.e.,
$\mathrm{Re}(-ig(1+\delta_{e})^{2}/\delta_{e}^{2})<0$. (d) Regions
satisfying all the (a), (b) and (c) realizability criteria. }
\end{figure}

\section{Backward wave oscillator (BWO) threshold current}

For a finite SWS with length $\ell$, the equivalent voltage that
describes the electric field at any coordinate $z$ in the SWS is
expanded as a combination of the four modes supported by the interactive
beam-EM mode system (modes are calculated in a SWS of infinite length)
as

\begin{equation}
V(z)=V_{1}e^{-ik_{1}z}+V_{2}e^{-ik_{2}z}+V_{3}e^{-ik_{3}z}+V_{4}e^{-ik_{4}z}\label{eq:Voltage_Dist}
\end{equation}
where $k_{n}$ with $n=1,2,3,4$ are the modes' wavenumber, and they
are determined from (\ref{eq:Disp_1}), and $V_{n}$ are the complex
modes amplitudes. The associated TL current and beam dynamics are
found by using (\ref{eq:Voltage_Dist}) in (\ref{eq:Sys_Equation_All})
or by using the eigenvector expression in (\ref{eq:eigenvector_1})

\begin{equation}
\begin{array}{c}
I(z)=\dfrac{ik_{1}V_{1}}{Z}e^{-ik_{1}z}+\dfrac{ik_{2}V_{2}}{Z}e^{-ik_{2}z}\ \ \ \ \ \ \ \ \ \ \ \ \ \ \ \\
\ \ \ \ \ \ \ \ \ \ \ \ \ \ \ \ \ \ \ \ \ +\dfrac{ik_{3}V_{3}}{Z}e^{-ik_{3}z}+\dfrac{ik_{4}V_{4}}{Z}e^{-ik_{4}z},\\
\\
V_{b}(z)=\dfrac{(1+\delta_{1})V_{1}}{\delta_{1}}e^{-ik_{1}z}+\dfrac{(1+\delta_{2})V_{2}}{\delta_{2}}e^{-ik_{2}z}\ \ \ \ \ \ \ \ \ \ \\
\ \ \ \ \ \ \ \ \ \ +\dfrac{(1+\delta_{3})V_{3}}{\delta_{3}}e^{-ik_{3}z}+\dfrac{(1+\delta_{4})V_{4}}{\delta_{4}}e^{-ik_{4}z},\\
\\
I_{b}(z)=\dfrac{g(1+\delta_{1})V_{1}}{\beta_{0}\delta_{1}^{2}}e^{-ik_{1}z}+\dfrac{g(1+\delta_{2})V_{2}}{\beta_{0}\delta_{2}^{2}}e^{-ik_{2}z}\ \ \ \ \ \ \ \ \ \ \ \ \ \\
\ \ \ \ \ \ \ \ \ \ \dfrac{g(1+\delta_{3})V_{3}}{\beta_{0}\delta_{3}^{2}}e^{-ik_{3}z}+\dfrac{g(1+\delta_{4})V_{4}}{\beta_{0}\delta_{4}^{2}}e^{-ik_{4}z},
\end{array}\label{eq:Voltage_Dist-1}
\end{equation}
where $\delta_{n}=(k_{n}-\beta_{0})/\beta_{0}.$ The setup we use
for the BWO with distributed power radiation, shown in Fig. 4(a),
is similar to that used in \cite{johnson1955backward}, where we assume
unmodulated space charge at the beginning of the electron beam (i.e.,
$V_{b}(z=0)=0$ and $I_{b}(z=0)=0$), and we also assume that the
output power is extracted at $z=0$, terminated with a resistance
matched to the characteristic impedance of the TL (without loss and
gain) $R_{o}=\sqrt{L/C}$, and short circuit at $z=\ell$, where $\ell=N\lambda_{e}$
is the SWS length (i.e the TL length), and $\lambda_{e}=2\pi/\beta_{0e}$
is the guided wavelength calculated at the EPD frequency. (Note that
in the absence of loss and gain in the TL, the TL distributed series
impedance and shunt admittance are $Z=1/(i\omega C)$ and $Y=1/(i\omega L)$).
The boundary conditions that describe the mentioned setup are

\begin{equation}
\begin{array}{c}
V(0)+R_{o}I(0)=0,\\
V(\ell)=0,\\
V_{b}(0)=0,\\
I_{b}(0)=0.
\end{array}\label{eq:BC}
\end{equation}

By imposing (\ref{eq:Voltage_Dist-1}) in (\ref{eq:BC}), we obtain
a homogeneous system of linear equations that is written in matrix
form as

\begin{equation}
\mathbf{A}(\omega,I_{0})\mathbf{V}=\mathbf{0}\label{eq:Gain_Oss}
\end{equation}
where $\mathbf{V}=[\begin{array}{cccc}
V_{1}, & V_{2}, & V_{3}, & V_{4}\end{array}]^{T}$. Free oscillation in the interactive system occurs when there is
a solution of (\ref{eq:Gain_Oss}) despite the absence of the source
term (the right hand side of (\ref{eq:Gain_Oss}) is equal to zero).
Therefore, oscillation occurs for a combination of radian frequency
and electron beam current that satisfy

\begin{equation}
\mathrm{det}(\mathbf{A}(\omega,I_{0}))=0.\label{eq:Osc_Cond}
\end{equation}

Since the solution of the above equation defines the threshold beam
current to start oscillations, such solution is denoted by $I_{0}=I_{th}$
and $\omega=\omega_{res}$ is the frequency of the oscillation.

\section{Asymptotic scaling of the EPD-BWO threshold beam current}

In this section we derive the oscillation condition and the asymptotic
scaling of the threshold beam current with the SWS length for the
proposed EPD-BWO assuming that the length of the structure tends to
infinity. We follow the traveling-wave tube theory used in \cite{pierce1947theoryTWT},
\cite{pierce1951waves} and \cite{johnson1955backward}, since the
synchronization involves mainly three waves, those with $Re(k_{n})>0$.
Assuming that these three wavenumbers have a slight variation of the
unperturbed beam wavenumber, i.e, $k_{n}\approx\beta_{0}$, we use
the eigenvector expression in (\ref{eq:eigenvector2}) to write the
TL circuit voltage and beam dynamics distribution along the $z$-direction
which are represented in terms of three modes as \cite{johnson1955backward}

\begin{equation}
\begin{array}{c}
V(z)=V_{1}e^{-ik_{1}z}+V_{2}e^{-ik_{2}z}+V_{3}e^{-ik_{3}z},\\
V_{b}(z)=\dfrac{V_{1}}{\delta_{1}}e^{-ik_{1}z}+\dfrac{V_{2}}{\delta_{2}}e^{-ik_{2}z}+\dfrac{V_{3}}{\delta_{3}}e^{-ik_{3}z},\\
I_{b}(z)=\dfrac{gV_{1}}{\beta_{0}\delta_{1}^{2}}e^{-ik_{1}z}+\dfrac{gV_{2}}{\beta_{0}\delta_{2}^{2}}e^{-ik_{2}z}+\dfrac{gV_{3}}{\beta_{0}\delta_{3}^{2}}e^{-ik_{3}z},
\end{array}\label{eq:Voltage_Dist-1-1-1}
\end{equation}
where $\delta_{n}=(k_{n}-\beta_{0})/\beta_{0}$. The wavenumbers $k_{n}$
with $n=1,2,3$ are with positive real part and two of them are with
positive imaginary part, let's say $n=1,2$ and the other one is with
negative imaginary part, $n=3$. A numerical example of the wavenumbers
is shown in Fig. 3. Here, we follow the same procedure used in \cite{johnson1955backward}
to obtain the oscillation condition which is based on imposing infinite
voltage gain $A_{v}=V(0)/V(\ell)\to\infty$ . By imposing the beam
boundary condition $V_{b}(0)=0$ and $I_{b}(0)=0$ in (\ref{eq:Voltage_Dist-1-1-1})
and after some mathematical manipulations the gain expression is written
in its simplest form as \cite{johnson1955backward}
\begin{equation}
\begin{array}{c}
A_{v}^{-1}e^{i\beta_{0}\ell}=\dfrac{e^{-i\beta_{0}\delta_{1}\ell}\delta_{1}^{2}}{\left(\delta_{1}-\delta_{2}\right)\left(\delta_{1}-\delta_{3}\right)}+\dfrac{e^{-i\beta\delta_{2}\ell}\delta_{2}^{2}}{\left(\delta_{2}-\delta_{3}\right)\left(\delta_{2}-\delta_{1}\right)}\\
\ \ \ \ \ \ \ \ \ \ \ \ \ \ \ \ \ \ \ \ \ \ \ \ \ \ \ \ \ \ \ \ \ \ \ \ \ \ \ +\dfrac{e^{-i\beta_{0}\delta_{3}\ell}\delta_{3}^{2}}{\left(\delta_{3}-\delta_{1}\right)\left(\delta_{3}-\delta_{2}\right)}=0.
\end{array}\label{eq:Gain_Exp}
\end{equation}
 We first neglect the term with $e^{-i\beta_{0}\delta_{3}\ell}$ in
(\ref{eq:Gain_Exp}) since we consider very large SWS length $\ell$
and know that $\mathrm{Im}(\delta_{3})<0$. Therefore the gain expression
reduces to

\begin{equation}
A_{v}^{-1}e^{i\beta_{0}\ell}\approx\dfrac{e^{-i\beta_{0}\delta_{1}\ell}\delta_{1}^{2}}{\left(\delta_{1}-\delta_{2}\right)\left(\delta_{1}-\delta_{3}\right)}+\dfrac{e^{-i\beta_{0}\delta_{2}\ell}\delta_{2}^{2}}{\left(\delta_{2}-\delta_{3}\right)\left(\delta_{2}-\delta_{1}\right)}.\label{eq:Gain_Exp-1}
\end{equation}

By defining $\Delta=(\delta_{1}-\delta_{2})$, hence $\delta_{1}=\delta_{a}+\Delta/2$
and $\delta_{2}=\delta_{a}-\Delta/2$ where $\delta_{a}=(\delta_{1}+\delta_{2})/2$
is the average, the gain expression is then written as

\begin{equation}
\begin{array}{c}
A_{v}^{-1}e^{i\beta_{0}\ell}\approx e^{-i\beta_{0}\delta_{a}\ell}\Bigg(\dfrac{\left(\delta_{a}+\Delta/2\right)^{2}e^{-i\beta_{0}\Delta\ell/2}}{(\Delta/2)\left(\delta_{a}-\delta_{3}+\Delta/2\right)}\\
\ \ \ \ \ \ \ \ \ \ \ \ \ \ \ \ \ \ \ \ \ \ \ \ \ \ \ +\dfrac{\left(\delta_{a}-\Delta/2\right)^{2}e^{i\beta_{0}\Delta\ell/2}}{(\Delta/2)\left(\delta_{a}-\delta_{3}-\Delta/2\right)}\Bigg),
\end{array}\label{eq:Gain_Exp-1-1}
\end{equation}

In close proximity of the EPD we know that two modes coalesce, i.e.,
$\delta_{1}\approx\delta_{2}$, thus we can assume that $|\Delta|\ll|\delta_{a}|,$
i.e., $\Delta\to0$ as we approach the EPD. It is important to point
out that although $\Delta$ is a very small value we can not neglect
its effect in the exponential function in (\ref{eq:Gain_Exp-1-1})
because we assume the SWS to be very long, however we can neglect
$\Delta$ in other places, i.e., we have $\delta_{a}\pm\Delta/2\approx\delta_{a}$
and $\delta_{a}-\delta_{3}\pm\Delta/2\approx\delta_{a}-\delta_{3}$.
Thus the gain expression finally reduces to Eq. (\ref{eq:Gain_Expr_Final}).
The first oscillation frequency occurs when the constraint on the
wavenumbers $k_{1}-k_{2}=2\pi/\ell$ is satisfied. A beam current
slightly away from the EPD one causes the wavenumbers to bifurcate
from the degenerate one $k_{e}$, following the Puiseux series approximation
\cite{welters2011explicit} as $k_{1}-k_{2}\approx-2\alpha\sqrt{I_{0}-I_{0e}}$,
where $\alpha$ is a constant. Therefore, the threshold beam current
is determined by solving $2\alpha\sqrt{I_{th}-I_{0e}}=-2\pi/\ell$
which yields the the asymptotic trend for the scaling of threshold
current in (\ref{eq:Ass_Scaling}).

This asymptotic trend is confirmed in Fig. 4 , where the threshold
current $I_{th}$ is calculated numerically by solving $\mathrm{det}(\mathbf{A}(\omega,I_{0}))=0$
for the beam current $I_{0}$, as discussed in Appendix C.\bibliographystyle{ieeetr}
\bibliography{myref}
 
\end{document}